\documentclass[a4paper,fleqn,usenatbib]{mnras}

\usepackage{newtxtext,newtxmath}
\usepackage[T1]{fontenc}
\usepackage{aecompl}
\usepackage{url}
\usepackage{graphicx}	
\usepackage{amsmath}	
\usepackage{amssymb}
\usepackage{appendix}

   \title[Characterisation of AGN from the XMM Slew Survey]{Characterisation of AGN from the {\it XMM-Newton} Slew Survey\thanks{Based on service observations made with the William Herschel Telescope operated on the island of La Palma by the Isaac Newton Group in the Spanish Observatorio del Roque de los Muchachos of the Instituto de Astrof\'isica de Canarias.}}

   \author[Starling et. al. 2016]
          {R. L. C. Starling$^{1}$, C. Wildy$^{1\thanks{Present address: The Center for Theoretical Physics of the Polish Academy of Sciences}}$, K. Wiersema$^{1}$, S. Mateos$^{2}$, R. D. Saxton$^{3}$,\and
            A. M. Read$^{1}$ and B. Mingo$^{1}$\\$^{1}$University of Leicester, Department of Physics and Astronomy, University Road, Leicester LE1 7RH, UK\\$^{2}$Instituto de F\'isica de Cantabria (CSIC-Universidad de Cantabria), 39005 Santander, Spain\\$^3${\it XMM-Newton} SOC, ESAC, Apartado 78, 28691 Villanueva de la Ca\~nada, Madrid, Spain}
            
\begin{document}
\date{}

\pagerange{\pageref{firstpage}--\pageref{lastpage}} \pubyear{2016}

\maketitle

\label{firstpage}
   \date{Received Month day, year; accepted}

  \begin{abstract}
    We present optical spectroscopy of candidate AGN pinpointed by a {\it Swift} follow-up campaign on unidentified transients in the {\it XMM-Newton} Slew Survey, increasing the completeness of the identifications of AGN in the Survey.
Our {\it Swift} follow-up campaign identified 17 XRT-detected candidate AGN, of which nine were selected for optical follow-up and a further two were confirmed as AGN elsewhere.  
    Using data obtained at the William Herschel Telescope, Very Large Telescope and New Technology Telescope, we find AGN features in seven of the candidates. We classify six as Seyfert types 1.0 to 1.5, with broad-line region velocities spanning 2000--12000 km s$^{-1}$, and identify one as a possible Type II AGN, consistent with the lack of a soft band X-ray detection in the Slew Survey. The Virial black hole mass estimates for the sample lie between 1$\times$10$^{8}$ M$_{\odot}$ and 3$\times$10$^9$ M$_{\odot}$, with one source likely emitting close to its Eddington rate, $L_{\rm Bol}/L_{\rm Edd} \sim 0.9$. We find a wide redshift range of $0.08<z<0.9$ for the nine now confirmed AGN drawn from the unidentified Slew Survey sample. One source remaining unclassified shows outbursts rarely seen before in AGN.  
    We conclude that AGN discovered in this way are consistent with the largely non-varying, Slew-selected, known AGN population. We also find parallels with {\it XMM-Newton} Bright Serendipitous Survey AGN selected from pointed observations, and postulate that shallow X-ray surveys select AGN drawn from the same populations that have been characterised in deeper X-ray selected samples.
\end{abstract}

  \begin{keywords}
    galaxies: active -- galaxies: Seyfert
\end{keywords}
%

\section{Introduction}
The XMM-Newton Slew Survey \citep{Saxton}
makes use of data taken while the {\it XMM-Newton} satellite is manoeuvering between pointed observations, reaching five to ten times deeper in flux than all other all-sky spatially-resolved surveys in the 2--12 keV band ($f_{\rm limit} \sim 3.7 \times 10 ^{-12}$ erg cm$^{-2}$ s$^{-1}$) and comparable sensitivity to the {\it ROSAT} PSPC All-Sky Survey \citep[RASS][]{Voges} in the 0.2--2 keV band ($f_{\rm limit} \sim 5.7 \times 10 ^{-13}$ erg cm$^{-2}$ s$^{-1}$).
The latest catalogue, XMMSL1$\_$Delta6 clean catalogue\footnote{\url{http://www.cosmos.esa.int/web/xmm-newton/xmmsl1d-ug}}, contains 20163 sources (17345 0.2-12\,keV full-band, 2160 2-12\,keV hard-band and 14371 0.2-2\,keV soft-band detections) covering 35350 square degrees. Among these are active galactic nuclei (AGN) \citep[e.g.][]{Miniutti,Strotjohann}, stellar systems \citep[e.g.][]{Torres,Lopez}, novae \citep[e.g.][]{Read2008,Read2009}, tidal disruption candidates \citep[e.g.][]{Esquej2007,Esquej2008,Saxton2012,Mainetti} and other transients.

A substantial fraction of XMM Slew Survey catalogued sources remain unclassified. In the hard, 2--10\,keV, band for example, \cite{Warwick} found that 37 per cent of Slew Survey sources (up to and including release XMMSL1d2, numbering $\sim$180) were still to be identified, while the latest Slew catalogue gives possible source identifications for all but 27 per cent of sources. \cite{Starling} (hereafter S11) took an unidentified XMMSL1$\_$Delta4 source sample, selected to contain sources detected in any of the three energy bands (hard 2--12\,keV, soft 0.2--2\,keV and full 0.2--12\,keV) with no counterpart within a 30 arcsecond radius in multiple-catalogue cross-matching, and followed these up with the {\it Swift} satellite \citep{Gehrels}. From the 94-strong {\it Swift}-observed sample, 29 per cent (27) were detected with the X-ray Telescope (XRT). For the first time, {\it Swift} allowed accurate astrometry for these sources in the X-rays, down to 1.5 arcseconds, leading to the identification of a single optical counterpart for most sources from considering both Swift UV-Optical Telescope data (positional accuracy 0.4$''$) and catalogue matches. For sources without a firm classification at this stage, this led to a list of candidate flare stars and candidate active galactic nuclei (AGN, see Table 8 of S11).

For AGN, selection via X-ray surveys is typically very efficient, picking up all but the most Compton thick of sources, and has the advantage of being unaffected by host galaxy contamination \citep{BrandtAlexander}. The Slew Survey is mapping bright X-ray sources, many of which are AGN, over a large fraction of the sky. To unambiguously classify these and determine their redshift distribution, optical spectroscopy is required to identify the characteristic broad and narrow AGN emission lines \citep[e.g.][]{Parisi,Saxton2014}.

The number of X-ray sources has steadily increased through surveys such as the XMM Serendipitous Source Survey \citep{Rosen} and the Swift X-Ray Telescope Point Source Catalog \citep{Evans2014}. In addition, X-ray tiling campaigns over the large error regions of transient events that include {\it Fermi} LAT gamma-ray bursts, neutrino events \citep[e.g.][]{Evans2015} and gravitational wave detections \citep[e.g.][]{Evans2016,Troja} is further increasing the need for initial characterisation and understanding of X-ray source populations. The planned four year all-sky survey with future instrument eROSITA \citep{Kolodzig} will reveal yet more of the X-ray sky, and will depend upon synergies with optical and IR spectroscopic facilities to fully exploit the rich datasets it is expected to provide \citep{Salvato}.

In this paper we present optical spectroscopy of eight of the candidate AGN drawn from the unidentified Slew sample, using the William Herschel Telescope (WHT), La Palma, and the Very Large Telescope (VLT), Chile. The primary aims were to confirm or refute the AGN nature of the sources and measure redshifts, which could be used to determine their physical properties such as nuclear absorption, radiative power and black hole mass. We also expand upon the analysis of our New Technology Telescope (NTT) data presented previously for one of these AGN (S11), and make use of published spectra from the Sloan Digital Sky Survey \citep[SDSS,][]{Alam} and the Large Quasar Astrometric Catalog \citep[LQAC][]{Souchay} and accompanying Multiple Mirror Telescope (MMT) spectrum \citep{Gioia}.
An understanding of the AGN types and redshift distribution present in this sample is important in the context of a census of black hole accretion across the Universe, and enhances the use of the wide-field XMM Slew Survey for AGN population and long-term variability studies.

The unidentified Slew AGN candidates and our optical spectroscopic observations are detailed in Section \ref{sec:datamethod} along with the modelling procedure. We interpret the spectral measurements in Section \ref{sec:results}. Section \ref{sec:nonagn} discusses the candidates which we find cannot be classified as Type I AGN. In Section \ref{sec:discuss} we compare the AGN we present with other samples of AGN drawn from {\it XMM-Newton} surveys and conclude.

\begin{table*}
     \caption{Full list of AGN and candidate AGN drawn from the XMM Slew Survey unidentified sample of S11, in order of increasing RA. WHT and VLT observations are presented here for the first time, while NTT observations presented in our previous work are explored in more detail here.}
         \label{tab:fullsample}
\begin{centering}
         \begin{tabular}{llll}
            Source name      & previously proposed ID (S11) & optical observations?&redshift  \\ \hline  
XMMSL1\,J002202.9+254004 & AGN/Blazar candidate&WHT&- \\  
XMMSL1\,J012240.2-570859 & AGN/NLS1 candidate&none&- \\
XMMSL1\,J030006.6-381617 & possible AGN candidate&VLT&-\\ 
XMMSL1\,J044357.4-364413 & possible AGN candidate&none&- \\
XMMSL1\,J064109.2-565542 & Type I AGN &NTT& $0.368\pm0.001$ \citep{Starling}\\ 
XMMSL1\,J065525.2+370815 & possible AGN/QSO candidate&WHT&-\\ 
XMMSL1\,J070846.2+554905 & possible AGN candidate& WHT&-\\ 
XMMSL1\,J095336.4+161231 & AGN/QSO & SDSS3 DR-12& $0.873055\pm0.000079$ \citep{Alam} \\ 
XMMSL1\,J125522.0-221035 & AGN/Blazar candidate&WHT&-\\ 
XMMSL1\,J131651.2-084915 & AGN candidate&WHT&-\\
XMMSL1\,J141843.5-293749 & AGN candidate&none &-\\
XMMSL1\,J162533.2+632411 & AGN/Type II candidate&WHT &-\\ 
XMMSL1\,J164859.4+800507 & possible AGN candidate&none &- \\
XMMSL1\,J175542.2+624903 & Type I AGN &MMT/LQAC& $0.236\pm0.001$ \citep{Gioia} \\
XMMSL1\,J182707.5-465626 & possible AGN candidate&none &- \\
XMMSL1\,J185608.5-430320 & possible AGN candidate&none &-  \\
XMMSL1\,J211420.7+252419 & possible AGN candidate&WHT &-\\ 
         \end{tabular}
\end{centering}
   \end{table*}
   
\section{Data and Method}\label{sec:datamethod}
\subsection{Sources} \label{sec:sources}
We considered all 17 sources from the unidentified Slew Survey sample presented in S11 which were detected by {\it Swift} XRT and listed as AGN or AGN candidates based on their X-ray properties, X-ray to optical flux ratios and 2MASS near-infrared colours (Table \ref{tab:fullsample}). Three sources from the list of 17 were already known to be AGN, with spectra observed previously by SDSS, LQAC and ESO NTT, leaving 14 AGN candidates. Seven of the AGN candidates were visible to WHT and we requested ACAM spectroscopy of these sources through a dedicated Service programme. The following section details these observations, alongside observations of a single AGN candidate we observed with the VLT as part of a poor-weather, back-up programme. The remaining 6 candidate AGN are not at suitable declinations to be included in our WHT programme and currently have no ground-based coverage that we are aware of.

\subsection{Observations}\label{sec:obs}
We obtained low resolution spectroscopy of seven AGN candidates from
the XMM Slew Survey with the auxiliary-port camera, ACAM, on the 4.2-m William Herschel Telescope (WHT), through Service
observing proposal Sw2011b12. Observations were carried out on three seperate nights: 2011-12-12 (hereafter, night 1),
2011-12-29 (night 2) and 2012-05-11 (night 3). Conditions were not spectro-photometric on all nights, precluding an
absolute flux calibration; this does not affect our ability to classify sources. We used a GG495 order-blocking filter and
the V400 volume phased holographic grating, resulting in a wavelength range $\sim$ 4950--9500\AA. Data were
reduced using standard techniques in IRAF.

We also obtained observations of one target with the FOcal Reducer and low dispersion Spectrograph 2 (FORS2) on the 8-m VLT. Using the standard star GD108, we were able to flux calibrate and apply a reddening correction \citep{Schlafly}.  

Observations with WHT and VLT are listed in Table \ref{tab:obs} and finding charts for all eight sources are shown in the Appendix in Figure \ref{fig:fc}. In the WHT and VLT observations we detected all eight sources, with emission lines visible in seven. 

   \begin{table}
      \caption{Log of spectroscopic observations taken with WHT and VLT. Those obtained at NTT are reported in S11.}
         \label{tab:obs}
\begin{centering}
         \begin{tabular}{lll}
            XMMSL1 source      &  Date-obs &  T$_{\rm exp}$ (s)\\ \hline
WHT/ACAM && \\ \hline
J002202.9+254004 & 2011-12-12& 5x300 \\
J065525.2+370815 & 2011-12-12& 3x1800 \\
J070846.2+554905 & 2011-12-29& 2x240,1x180 \\
J125522.0-221035 & 2011-12-29& 2x1500 \\
J131651.2-084915 & 2011-12-29& 2x900 \\
J162533.2+632411 &2011-12-29& 2x1500 \\
J211420.7+252419 &2012-05-11 &2x180  \\ \hline
VLT/FORS2&& \\ \hline
J030006.6-381617 & 2012-03-01 &1x600,1x300 \\ \hline
    \end{tabular}
\end{centering}
   \end{table}

\begin{table*}
\begin{centering}
\caption{Results of optical line fitting. All full width half maxima (FWHM) are given in km s$^{-1}$. Offsets compare the central positions of the total broad component with the narrow component in km s$^{-1}$. The doublets of forbidden [O\,III], [O\,I] and [N\,II] were locked and we therefore report the fit for one component only. $^*$Linewidth given is for the whole H$\beta$ blend. $^+$A single, weak, likely narrow H$\alpha$ emission line is detected, but cannot be modelled to obtain a linewidth due to line blending.}
\begin{tabular}{l| l| l| l| l| l| l| l}
  & \multicolumn{7}{c}{Source name (abbreviated)}\\
 & J002202 & J030006 & J064109 & J125522 & J131651 &J162533& J211420 \\\hline \hline
H$\beta$: no. broad components&2&2&-&1&1&-&1\\
H$\beta$: total broad FWHM& 8900$^{+300}_{-400}$&4300$^{+400}_{-100}$&$^*$9300$^{+1900}_{-600}$&10100$\pm$300&3900$\pm$400&-&3500$\pm$100\\
H$\beta$: narrow FWHM&1200$\pm$100 &600$^{+100}_{-200}$&-&2000$\pm$300&1100$\pm$100&-&1200$\pm$100\\
H$\beta$: offset&$+$1300& $+$74& - & $+$840&$+$1100&-& $-$110 \\
Fe\,II$\lambda$4923\AA\ FWHM&2500$^{+500}_{-600}$&-&-&2700$^{+2000}_{-1400}$&-&-\\
Fe\,II$\rbrack\lambda$4928\AA\ FWHM&-&-&-&-&-&-&1900$\pm$100\\
$\lbrack$O\,III$\rbrack\lambda$5007\AA\ FWHM&1000$\pm$100&500$\pm$100&900$\pm$100&1100$\pm$100&800$\pm$100&-&700$\pm$100\\
Fe\,II$\lambda$5018\AA\ FWHM&-&-&-&-&5000$^{+600}_{-2300}$&-&3700$\pm$300\\
$\lbrack$Fe\,II$\rbrack\lambda$5020\AA\ FWHM&-&-&-&-&-&-&1300$^{+100}_{-300}$\\
$\lbrack$O\,I$\rbrack\lambda$6300\AA\ FWHM &800$\pm$100&700$\pm$100&-&900$\pm$800&-&-&-\\
$\lbrack$N\,II$\rbrack\lambda$6583\AA\ FWHM &1000$\pm$100&400$^{+200}_{-100}$&-&-&600$\pm$100&-&1800$^{+100}_{-400}$\\\hline
H$\alpha$: no. broad components&3&2&-&1&1&1$^+$&1\\
H$\alpha$: total broad FWHM&7900$\pm$100&3900$\pm$100&-&9300$^{+100}_{-300}$&3300$\pm$200&-&3000$\pm$100\\
H$\alpha$: narrow FWHM&900$\pm$100&500$\pm$100&-&1900$^{+100}_{-300}$&900$\pm$100&-&1200$\pm$100\\
H$\alpha$: offset&$+$510&$-$44&-&$-$750&$+$120&-&$-$590\\ \hline 
{\bf Redshift, $z$} &{\bf 0.12920}&{\bf 0.2467}&{\bf 0.3684}& {\bf 0.3391}&{\bf 0.1381}&{\bf 0.104}$^+$&{\bf 0.0892} \\
$z$ error & $\pm$0.00004& $\pm$0.0001&$\pm$0.0002&$\pm$0.0009&$\pm$0.0001&&$\pm$0.0002 \\\hline
H$\beta$/[O\,III]&2.35&1.55& 1.44& 6.83&0.79&-&5.80\\
{\bf Seyfert subclass}&{\bf Sy 1.2}& {\bf Sy 1.5}& {\bf Sy 1.5}&{\bf Sy 1.0}& {\bf Sy 1.5}& {\bf Type II?}&{\bf Sy 1.0} \\ \hline
\end{tabular}
\label{tab:lines}
\end{centering}
\end{table*}

\begin{figure*}
\begin{centering}
\includegraphics[width=0.45\textwidth]{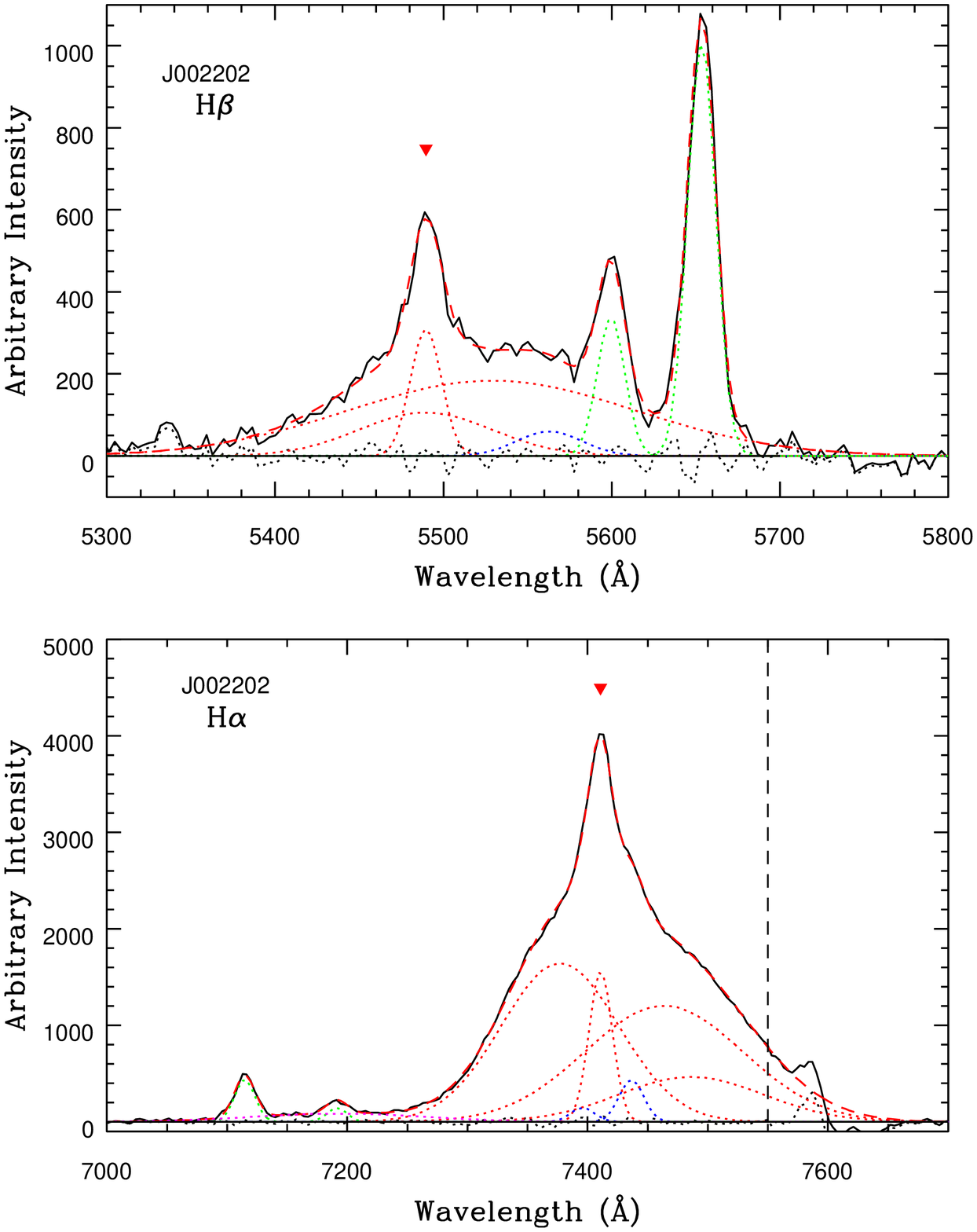}
\includegraphics[width=0.45\textwidth]{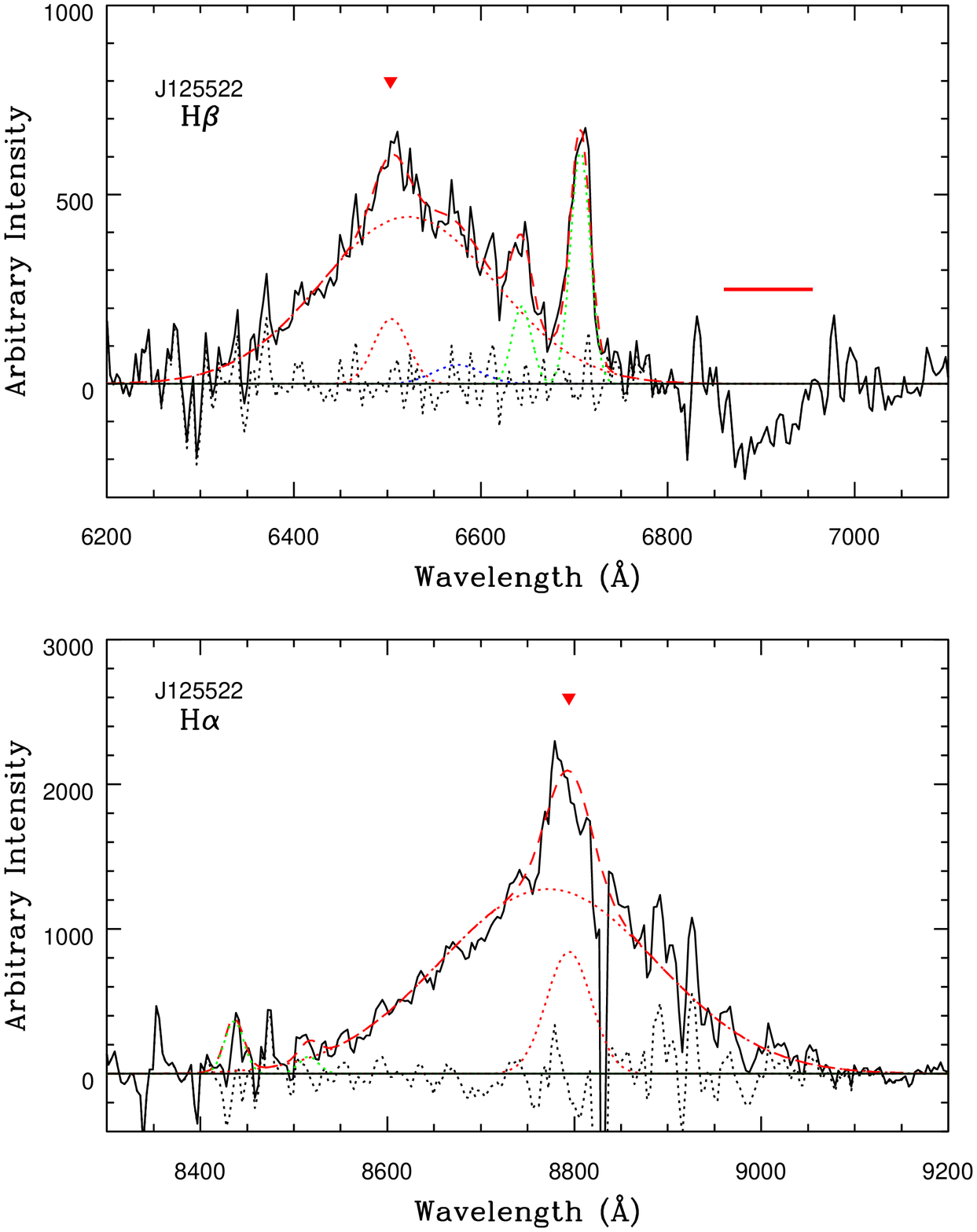}
 \includegraphics[width=0.45\textwidth]{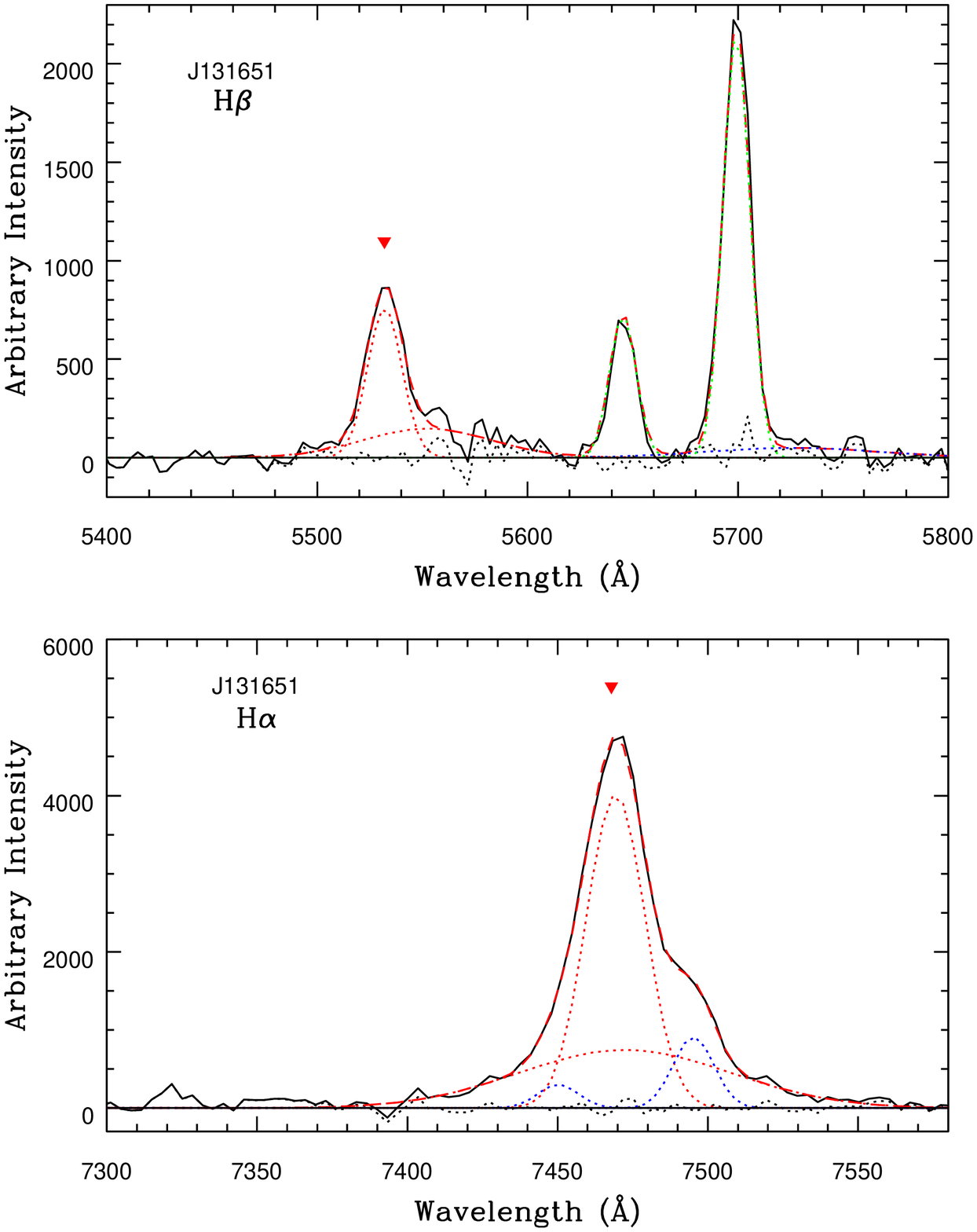}
\includegraphics[width=0.45\textwidth]{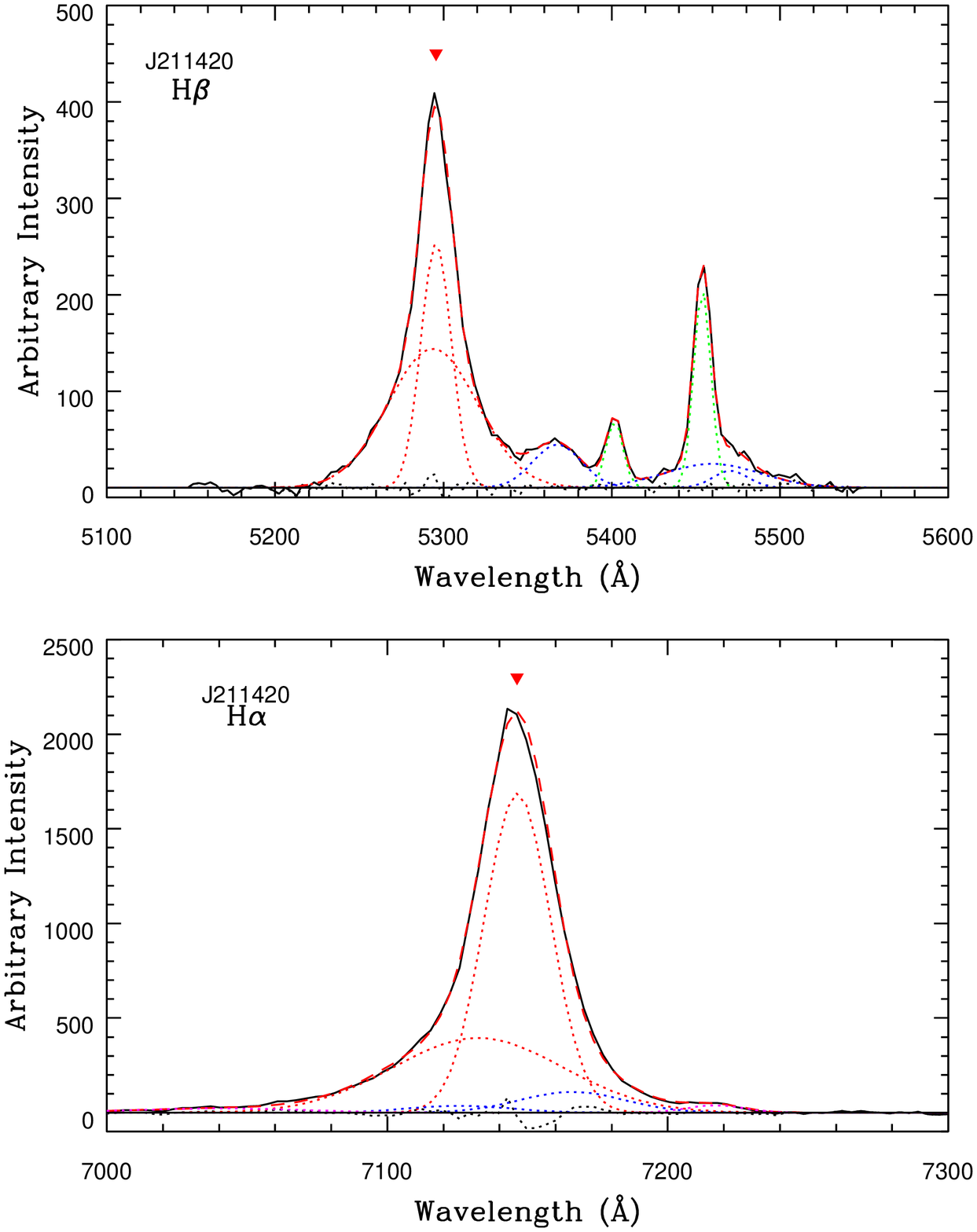}
\caption{Optical spectral line fits with local continuum subtracted: a zoom-in of the H$\alpha$ and H$\beta$ regions for the candidate AGN observed with WHT. Colour/symbol key: solid black line = observed spectrum; dashed red line = total model; dotted black line = residuals; dotted red line = Balmer line components; red triangle = Balmer narrow-line centroid; dotted green line = [O\,III](H$\beta$ blend), [O\,I](H$\alpha$ blend); dotted blue line = Fe\,II (H$\beta$ blend), [N\,II](H$\alpha$ blend); dashed black vertical line = no fitting redward due to sky absorption feature; red horizontal bar = telluric feature.}
\label{fig:linefitsall}
\end{centering}
\end{figure*}

\begin{figure}
\begin{centering}
\includegraphics[width=0.45\textwidth]{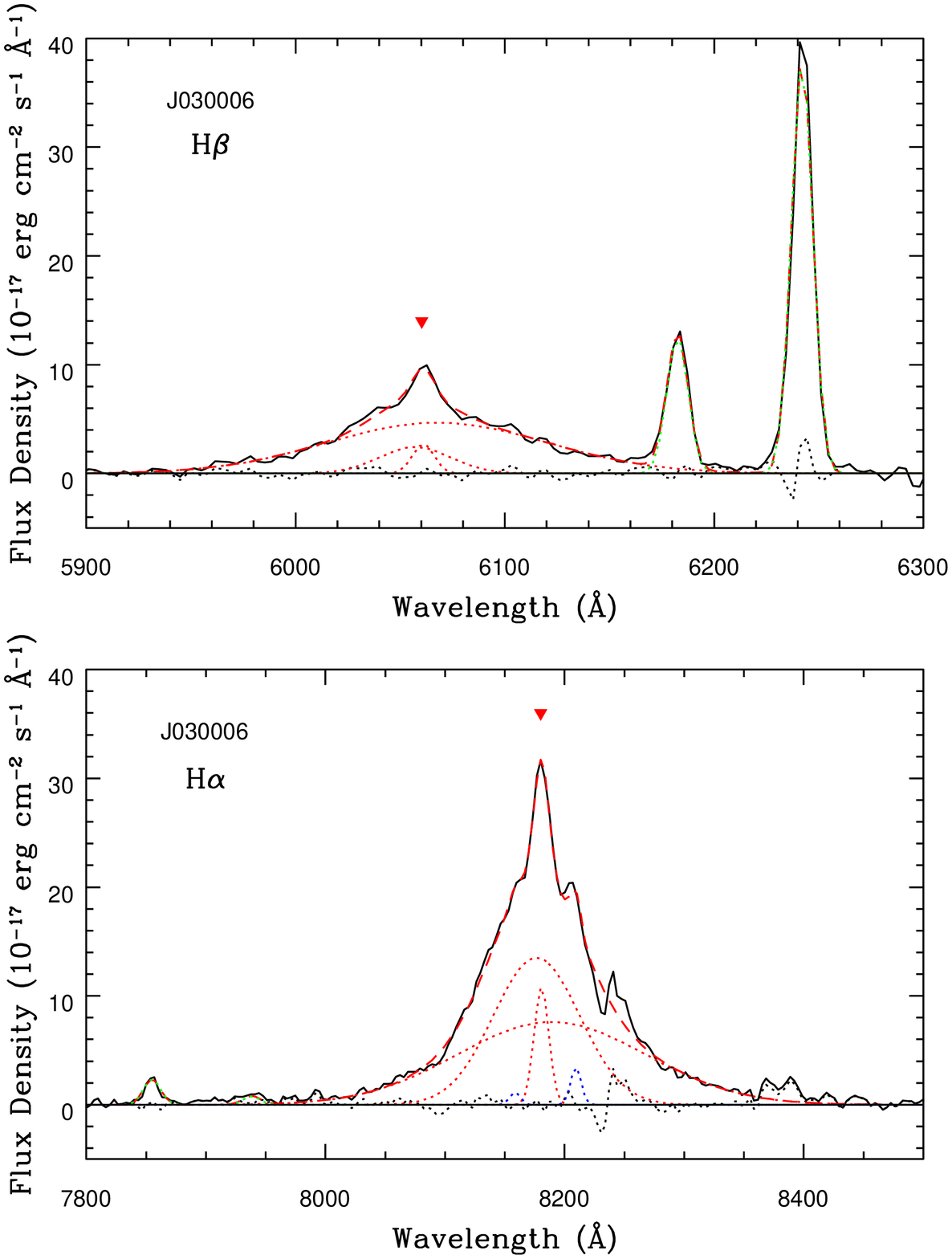}
\includegraphics[trim={0 12cm 0 0},clip=true,width=0.45\textwidth]{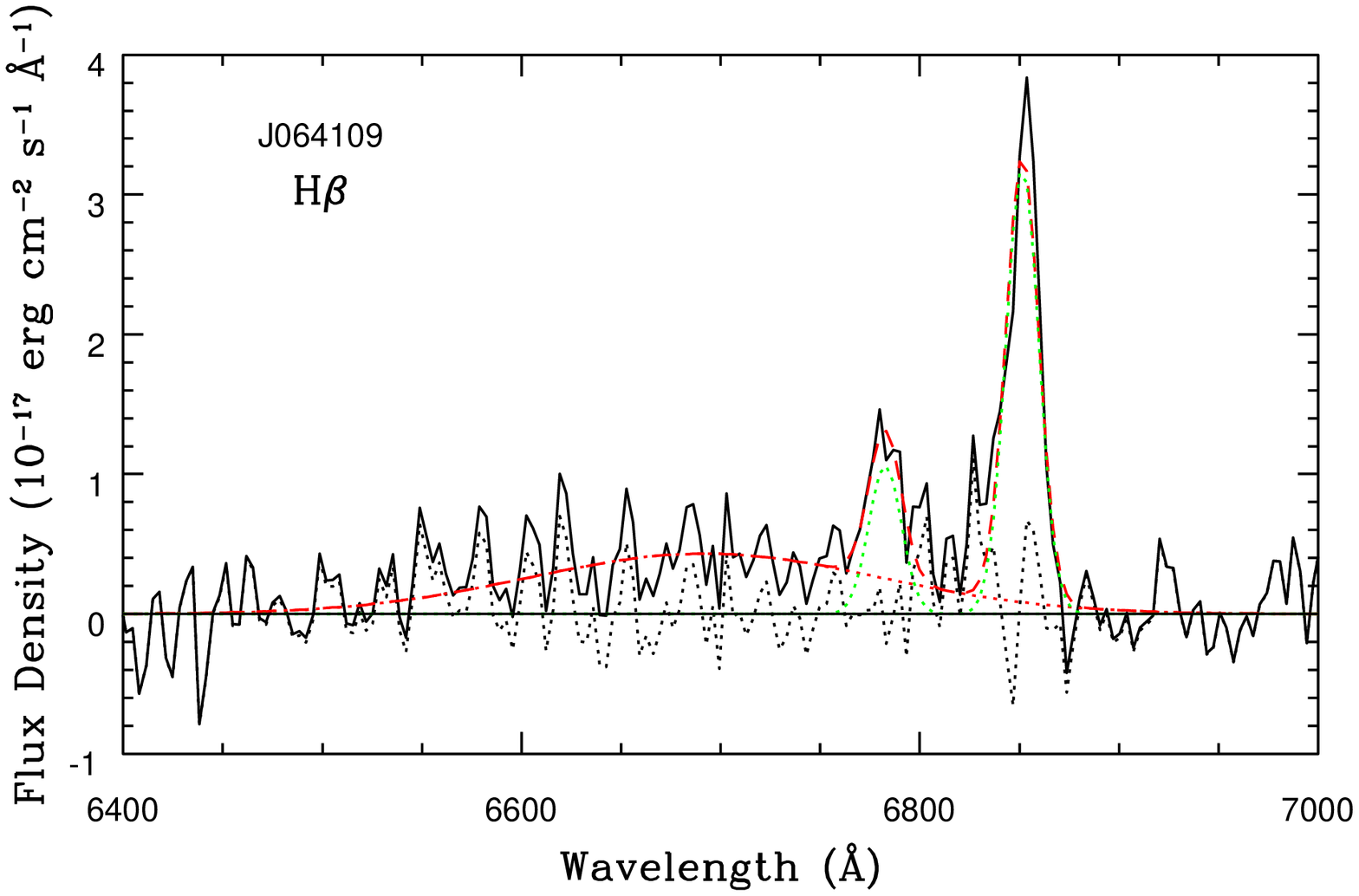}
\caption{Optical spectral line fits with local continuum subtracted: a zoom-in of the H$\alpha$ and H$\beta$ regions for the candidate AGN observed with VLT and NTT. Colour/symbol key identical to Figure 1.}
\label{fig:linefitsVLTNTT}
\end{centering}
\end{figure}

\subsection{Optical spectral line fitting}\label{sec:lines}
In our WHT and VLT observations broad and narrow redshifted emission lines were present in five sources, and in a further source we detect a single, weak, redshifted line. We added to these six a further source from the full sample (Table \ref{tab:fullsample}), XMMSL1\,J064109.2-565542, whose NTT EFOSC2 observation was presented previously and covers the H$\beta$ region only, but detailed line fits were not carried out (see S11). Spectral fitting was performed on the seven candidate AGN\footnote{A redshift was reported for a further candidate, XMMSL1\,J175542.2+624903, in \cite{Gioia} and listed in our Table \ref{tab:fullsample}, but spectral line measurements are not reported and unfortunately those data are no longer available.} using the specfit package within IRAF. We fitted a local continuum around each cluster of emission lines, and modelled the lines with one Gaussian for narrow components and up to three Gaussians for broad components as required. This enabled us to measure line widths and distinguish Type I from Type II AGN. For each source the modelling provided the peak position of the major narrow emission lines from which we took an average to determine the AGN redshift. Furthermore, we obtained line ratios (these are approximate in the absence of flux calibration and reddening-correction) which could be used to further classify our sources. A summary of line fitting results and properties derived from these fits is given in Table \ref{tab:lines} and shown in Figures \ref{fig:linefitsall} and \ref{fig:linefitsVLTNTT}.

\section{Results}\label{sec:results}
\subsection{Spectral classifications and redshifts}\label{sec:class}
Five sources showed multiple well-detected emission lines, and one source showed a broad H$\beta$ line to which we could fit profile models (Figures \ref{fig:linefitsall}, \ref{fig:linefitsVLTNTT}). Where more than one Gaussian was required to fit the broad component of an emission line, these were combined to give a total broad linewidth. Their broad line widths have velocities of 2000--12000 km s$^{-1}$ full-width at half maxima (FWHM), and we classify all of these as Type I AGN. 

We can assign a Seyfert subclass to each source using the H$\beta$/[O\,III] line ratio, where a value greater than 5.0 indicates a Seyfert 1.0, 2.0--5.0 indicates a Seyfert 1.2, 0.3--1.9 indicates a Seyfert 1.5 \citep[][and references therein]{Winkler}. These lines are sufficiently close in wavelength that flux calibration should result in a negligibly small difference in line ratio compared with the value reported here. The sources span Seyfert 1.0--1.5 subclasses - all Type I AGN (Table \ref{tab:lines}). No narrow-lined Seyferts are found, using the criterion of \citet{Osterbrock} of H$\beta$ FWHM $<$2000 km s$^{-1}$. 
The AGN from the unidentified Slew sample which were confirmed elsewhere, XMMSL1\,J095336.4+161231 and XMMSL1\,J175542.2+624903, are also of Type I. 

With our WHT observations we also detected XMMSL1\,J162533.2+632411, a galaxy with weak H$\alpha$ emission, and weak Na\,I and Mg\,Ib absorption complexes. We were unable to perform profile fitting on the H$\alpha$ line, but were able to obtain a redshift. We tentatively classify this source as a Type II, highly absorbed AGN on account of the narrow appearance of H$\alpha$ and lack of a soft X-ray band detection in the Slew Survey, and include it in Table \ref{tab:lines}.

Furthermore, we detected XMMSL1\,J070846.2+554905 with WHT. We find this to be a Galactic object, likely stellar in nature. We detect several stellar absorption
lines at $z=0$ (including the Ca\,II triplet, Mg\,Ib and Na\,I) and a weak H$\alpha$ line in emission. We will not discuss this source further, and it is not included in Table \ref{tab:lines}.

Our WHT observation of XMMSL1\,J065525.2+370815 is inconclusive. The faintness of the source in catalogue image servers required acquisition using an offset star. In the 2D spectrum a weak, barely significant, trace is seen. No absorption or emission features were detected from this object, though
the signal to noise is too low to rule out any classification for this source.

In order to obtain a redshift measurement for each emission-line source, we averaged the best-fitting peak positions from the narrow lines of H$\alpha$, H$\beta$ and [O\,III]. The redshift errors were calculated using the root-mean-square of the deviations of these positions from the mean. The resulting redshift range spans $0.08<z<0.25$. Adding in the previously known redshifts increases the redshift range to $0.08<z<0.9$.

\subsection{Black hole mass estimates}\label{subsec:mass}
We estimated the black hole masses of all the AGN we confirm here (Table \ref{tab:lines}) and the previously known AGN from Table \ref{tab:fullsample} with broad line measurements, using the empirical relationship between broad-line region size and monochromatic luminosity at 5100\AA\
and applying the Virial Theorem.
This uses the newly obtained redshifts and H$\beta$ linewidths\footnote{The H$\beta$ linewidth for XMMSL1\,J095336.4+161231 comes from a published single Gaussian$+$background fit to the SDSS3 DR12 spectrum (plate 2583, fiber 0109, \citealt{Alam}) with $\sigma=1466$ km s$^{-1}$, from which we calculate $FWHM = 2 \sqrt{2 ln 2}\, \sigma$.} and catalogued optical photometry from USNOB1.0 \citep{Monet} and NOMAD \citep{Zacharias}. We note that the H$\beta$ linewidths are slightly larger than the H$\alpha$ linewidths for most of our sources, which is in agreement with previous studies \citep[e.g.][]{Greene}.

We follow the method outlined in \citet{Shen}, with coefficients taken from \citet{Vestergaard}:
\begin{equation}
log \left( \frac{M_{\rm BH}}{M_{\odot}} \right) = 0.672 + 0.61\,log \left( \frac{\lambda L_{\lambda}}{10^{44} erg~s^{-1}} \right) + 2\,log \left( \frac{FWHM}{km~s^{-1}} \right)
\end{equation}

We caution that this method only provides order of magnitude estimates for the black hole mass. We do not account for the effects of variability in 5100\AA\ fluxes or for stellar contamination by the host galaxies. The results we obtain span 7$\times~$10$^{7}$ -- 2$\times$10$^{9}$ M$_{\odot}$.

   \begin{table*}
     \caption{Estimates of black hole mass and Eddington ratio. Optical photometry from USNOB1.0 (R,I) or NOMAD (V) is given for the observed band closest to restframe 5100\AA. Black hole mass estimates use the Virial method described in \citet{Shen,Vestergaard} ($^*$H$\beta$ linewidth taken from SDSS for this source). Bolometric luminosity estimates, $L_{\rm Bol}$, use the corrections given in equation 21 of \citet{Marconi2004}.} 
         \label{tab:bhmass}
\begin{centering}
         \begin{tabular}{llllllll}
           XMMSL1 source & $\frac{5100\rm{\mathring{A}}}{1+z}$ & $m_{5100\rm{\mathring{A}},rest}$& broad H$\beta$ &$M_{\rm BH}$ &$L_{2-10\,keV}$&$L_{\rm Bol}$  & $L_{\rm Bol}/L_{\rm Edd}$  \\
      & & &  FWHM (km s$^{-1}$) & (M$_{\odot}$)& (10$^{44}$ erg s$^{-1}$)& (10$^{45}$ erg s$^{-1}$) &  \\\hline
J002202.9+254004 & 5758&V=17.96& 8900$^{+300}_{-400}$&4.7$\times$10$^{8}$&1.9$\pm$0.2&7.9&0.12\\ 
J030006.6-381617 & 6360&R=19.71&4300$^{+400}_{-100}$ & 7.3$\times$10$^{7}$&0.7$\pm$0.2&2.0&0.19\\ 
J064109.2-565542 & 6977 & R=17.96 & 9300$^{+1900}_{-600}$& 1.2$\times$10$^{9}$&5.4$^{+2.1}_{-1.5}$&26.8&0.16\\    
J095336.4+161231 & 9537 &I=16.52 &3452$^*$ &6.3$\times$10$^{8}$&11.6$^{+3.6}_{-3.0}$&81.1&0.93\\ 
J125522.0-221035 & 6829 &R=17.30&10100$\pm$300&1.8$\times$10$^{9}$&4.0$^{+2.5}_{-1.3}$&18.9&0.08\\ 
J131651.2-084915 & 5804&V=17.25&3900$\pm$400&1.3$\times$10$^{8}$&0.05$^{+0.02}_{-0.01}$&0.08&0.004\\ 
J162533.2+632411 &5630 &V=17.97&-&- &0.10$^{+0.06}_{-0.04}$&0.2&- \\
J175542.2+624903 & 6304&  R=16.28&-& -&1.0$\pm$0.2& 3.3&-\\ 
J211420.7+252419 & 5549 &V=15.50&3500$\pm$100&1.5$\times$10$^{8}$&0.25$^{+0.08}_{-0.06}$&0.9&0.04\\ 
\end{tabular}
\end{centering}
\end{table*}
   
\subsection{X-ray luminosities and Eddington fractions}\label{subsec:lumins}
We returned to the {\it Swift} X-Ray Telescope (XRT) X-ray spectra we presented in S11 in light of the new redshift measurements, and fitted each spectrum with an absorbed power law model in {\sc XSPEC}, making use of the most up-to-date Galactic column densities \citep{Willingale} and adopting Cash statistics. {\it Swift} XRT observations additional to our fill-in programme were available for one source, XMMSL1\,J002202.9+254004. We included these in a combined X-ray spectrum totalling 12\,ks of exposure, created using the method of \cite{Evans2009} and after establishing no significant changes in spectral hardness had occurred. 

From the XRT best fits we obtained the observed 2--10\,keV X-ray luminosities of our AGN. We estimated the Eddington luminosities, $L_{\rm Edd}$, using our mass estimates in the equation $L_{\rm Edd} = 1.38 \times 10^{38} M/M_{\odot}$\,erg\,s$^{-1}$.
To obtain the Eddington ratio, the bolometric luminosity is required. Our measured X-ray luminosities significantly underestimate the total output, by an approximate factor of ten \citep{Elvis}. \cite{Marconi2004} derived a bolometric correction to convert from 2--10 keV X-ray luminosity to $L_{\rm Bol}$ (their equation 21), which we adopt here. The resulting bolometric luminosities and Eddington ratios are listed in Table \ref{tab:bhmass}.

Most of these AGN have X-ray luminosities which imply a bolometric luminosity around a tenth of a percent Eddington (mean value 0.2). The spread is large however; our least luminous AGN, XMMSL1\,J131651.2-084915 with $L_{2-10keV} = 5 \times 10^{42}$ erg s$^{-1}$, is emitting at a few hundredths of the Eddington rate, whilst our most distant source, XMMSL1\,J095336.4+161231 at $z=0.873$, may be emitting at or close to the Eddington rate.

\subsection{Gamma-ray and radio searches}

At the hard X-ray/$\gamma$-ray energies of 15--150\,keV none of our sources are detected in the {\it Swift} BAT transient monitoring programme spanning 2005 to May 2016 (H. Krimm, private communication). This supersedes the low significance detections we tentatively reported in S11.

All of our sources, listed in Table \ref{tab:fullsample}, have coverage in either the NRAO VLA Sky Survey (NVSS, \citealt{Condon}) or Sydney University Molonglo Sky Survey (SUMSS, \citealt{Mauch}), three are also covered by the VLA Faint Images of the Radio Sky at Twenty centimeters survey (FIRST, \citealt{White}) and one by the Westerbork Northern Sky Survey (WENSS, \citealt{Rengelink}). A search in radio catalogues at these positions results in no spatially coincident radio sources (within 30$''$ radius). However, XMMSL1\,J164859.4+800507 (for which there is no optical spectroscopy) may be associated with a cluster of galaxies at $z < 0.25$ which contains the WENSS radio source WNB1652.5$+$8009 \citep{Edge}, located 33$''$ from the XMMSL1 source.

\section{Sources with non-Seyfert-1 classifications} \label{sec:nonagn}

\begin{figure}
  \begin{centering}
    \includegraphics[width=0.45\textwidth]{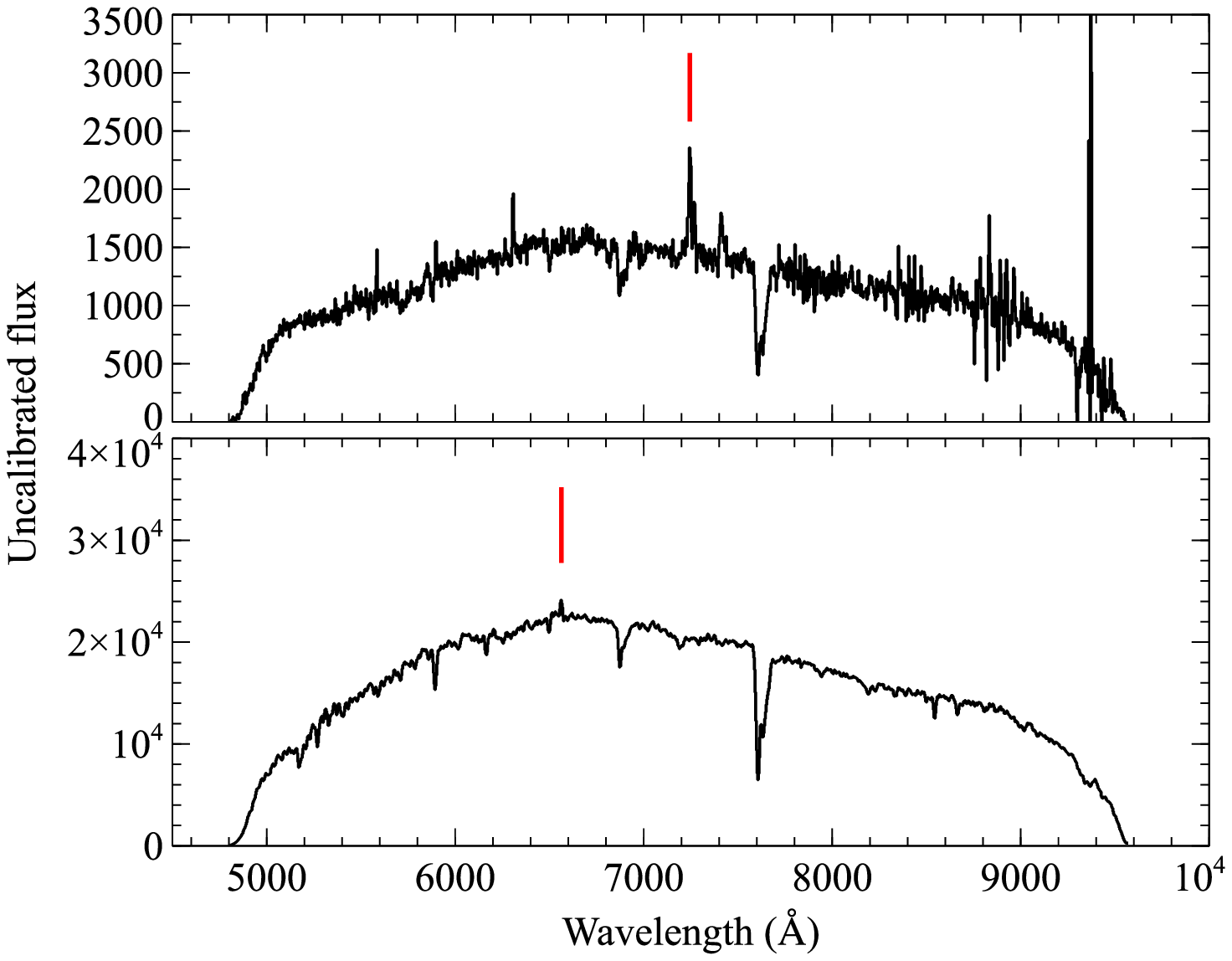}
\caption{Spectra obtained for XMMSL1\,J162533.2+632411 (upper plot) and XMMSL1\,J070846.2+554905 (lower plot), shown in instrumental fluxes (i.e. not corrected for instrumental response). The position of H$\alpha$ is indicated, and lies at $z = 0$ in the case of J070846 and $z = 0.104$ in the case of J162533.}
\label{fig:otherspectra}
\end{centering}
\end{figure}

\subsection{XMMSL1\,J070846.2+554905: a stellar object}
XMMSL1\,J070846.2+554905 was a bright X-ray source when observed with {\it Swift} but whilst its X-ray to optical flux was clearly AGN-like, its nIR colours indicated consistency with main sequence K stars. 
The spectrum we obtain, shown in Fig. \ref{fig:otherspectra}, confirms this as a stellar source, showing several stellar absorption lines and weak H$\alpha$ emission at $z=0$.

\subsection{XMMSL1\,J065525.2+370815: a highly unusual outbursting source or TDE?}
XMMSL1\,J065525.2+370815 is a highly variable X-ray source, which changed in flux by at least a factor 20 between the Slew and {\it Swift} observations. Three magnitudes of variability were detected between the catalogued B magnitude and the UVOT b band observation, and the UVOT data are correlated with the XRT fluxes. The SDSS image, taken coincident with the observed X-ray peak in 2006, shows an apparently extended source with a blue colour. A weak trace is visible in the 2011 WHT ACAM imaging, although this is not sufficient to extract a spectrum. We therefore either find this source in its low flux state at this time, or we are detecting the faint host galaxy of a now faded transient.

We analysed the catalogued AllWISE colours of this source \citep[obtained from a combination of observations from 2010 onwards][]{Cutri} in the context of the WISE colour-colour diagram for XMM-selected AGN samples \citep[see e.g.][]{Mateos2012,Mateos2013,Mingo}. The colours of this source are not stellar and neither do they lie in the region occupied by typical AGN. It appears to lie in the overlap region between star forming galaxies and luminous red galaxies. This supports the idea that we are observing a galaxy, superimposed upon which is a transient event.

The X-ray light curve shows a single peak with subsequent fading, characterised by two detections in 2003 and 2006 and two upper limits in 2008 and 2016, which are broadly consistent with the t$^{-5/3}$ decay expected from tidal disruption events. In the XRT detection we see a soft but poorly constrained spectrum ($\Gamma \sim 2.5$, S11). We searched the Catalina Sky Survey\footnote{\url{http://www.lpl.arizona.edu/css/}} data and found optical V-band observations of this source which show a strong, prolonged (at least $\sim$ year-long) brightening in 2005--2006, but also show a second rebrightening in 2010, reaching a similar peak magnitude but with a shorter time spent at peak flux, before returning to the quiescent level. This is more suggestive of a Galactic binary event, such as a recurrent nova, but could also indicate an AGN accretion disk instability as proposed for flares with decade-timescale separations in IC3599 \citep{Grupe2015}. 
More exotic explanations for the double outburst could include tidal stripping of an evolved star on two seperate encounters \citep{Mandel,Campana2015}, tidal disruption of a binary system \citep{Mandel}, or tidal disruption of a star by a black hole binary \citep{Coughlin}.

The lack of coverage, both in terms of wavelength and of sampling, unfortunately limits the insight we can gain into this nature of this source at present. We are therefore unable to classify XMMSL1\,J065525.2+370815. 

\subsection{XMMSL1\,J162533.2+632411: an obscured AGN}
XMMSL1\,J162533.2+632411 was designated a quasar candidate in automated classification analysis of SDSS DR10 objects \citep{Brescia}. We detected the source with the WHT but we found only very weak lines with which to estimate a redshift of $z=0.104$ (Fig. \ref{fig:otherspectra}). The source is remarkably hard in the X-rays, having both a hard band and a full band detection in the Slew Survey - the only one of our candidate AGN not present in the Slew soft band. While counts above 2 keV are seen in the XRT spectrum, the source remained very hard, with a hardness ratio of 2--10\,keV/0.3--2\,keV $=$ 3.5. It could be highly absorbed, and this must then be intrinsic absorption as the Galactic X-ray column density towards the source is low ($N_{\rm H} = 1.98 \times 10^{20}$ cm$^{-2}$, \citealt{Willingale}). The WHT spectrum is consistent with this.

Simple absorbed power law fits to the 0.3--10\,keV XRT spectrum are unable to constrain the intrinsic absorbing column density (S11), which is unsurprising given the low number of counts ($\sim20$). If, instead, we fix the hardness ratio, the observed flux and the Galactic column density, and we assume a power law photon index of $\Gamma=1.9$, we retrieve an X-ray column density at $z=0.104$ of $\sim$6$\times$10$^{21}$ cm$^{-2}$ which, while not large as is typical of Seyfert 2 galaxies \citep[e.g.][but note there are low $N_{\rm H}$ exceptions]{Risaliti,Mateos2005}, is certainly comparable to that observed in Seyferts of type 1--1.5 \citep[e.g.][]{Mateos2010,Ricci}. We note this fit provides a single estimate, assuming specific spectral parameters, while variability in column density is observed in Seyferts \citep[e.g.][]{MingoMrk6,Starling2014}.

Notably, the best available X-ray source location, from {\it Swift} XRT, lies within the 2$\sigma$ error circle of RASS source 1RXS\,J162535.1+632333 from the faint source catalogue \citep{Voges}. The position centroids are 26 arcseconds apart, and the count rate measured in the RASS was similar to that measured in the equivalent band in the XRT spectrum (0.3--2 keV XRT/RASS flux ratio = 1.3, using WebPIMMs to convert from {\it ROSAT} count rate to flux).
We examined its infrared output using the AllWISE magnitudes \citep{Cutri}. While its mid-IR colours locate the source in the AGN/QSO locus of the WISE colour--colour diagram for XMM-selected AGN samples \citep[see e.g.][]{Mateos2012,Mateos2013,Mingo}, and both the W1-W4 diagnostic ($\sim$ 6.6, see \citealt{Rovilos}) and the $L_{\rm X,2-12 keV}/L_{\rm 12 \mu m}$ ratio ($\sim$ 1, see \citealt{Gandhi,Rovilos,Asmus,Mingo}) indicate that the source is unlikely to fall in the heavily absorbed regime, it seems likely that it has some intrinsic obscuration, $\sim 10^{22}$ cm$^{-2}$, consistent with this source being a Type II or, given the soft X-ray variability, an intermediate Type I--II, rather than a Type I in line with the rest of our sample. 

\section{Discussion}\label{sec:discuss}
The XMM Newton Slew Survey is proving a useful resource for cataloguing X-ray objects, owing to its large sky coverage. A significant fraction of these objects do not have identifications, and a sample of 94 of these were followed up with {\it Swift}, of which 27 were detected with XRT opening up the possibility of source classification (S11). 63 per cent (17/27) of XRT detections were candidate AGN, of which 2 could be confirmed from catalogue matches following the improved astrometry from {\it Swift} observations. Our optical spectroscopy at WHT, VLT and previously NTT, has confirmed an AGN nature for 6 sources and strongly suggests an AGN nature for a further source, while a further source is found to be Galactic. There are 6 sources remaining for which optical spectroscopy has not yet been attempted, due to a lack of suitable facilities with service observing at certain declinations. 65 per cent (11/17) of the candidate AGN have now been followed up, with 9 out of 11 definitively shown to be AGN.

Most of our confirmed AGN are of Type I (89 per cent). This is expected, as X-ray bright AGN samples are often dominated by broad-lined AGN (e.g. 70 per cent in the XXL-BOSS sample, \citealt{Menzel}, and 88 per cent in the XBS sample, \citealt{DellaCeca,Caccianiga}). The average redshift of all our confirmed AGN is $z=0.28$,
average black hole mass is $M_{\rm BH} = 5.5 \times 10^8$ M$_{\odot}$ and average Eddington ratio is $L/L_{\rm Edd} = 0.2$, all with large associated standard deviations.

The implied black hole mass and bolometric luminosity of one of our sources, XMMSL1\,J095336.4+161231, suggest it could be radiating close to the Eddington limit. The X-ray flux of this AGN is well constrained in the combined XRT spectrum, but is variable over the two observations by a factor of 1.1--3.0 in 16 months (with no change in hardness ratio) and is not contemporaneous with the pre-2003 USNO-B1.0 \citep{Monet} I-band magnitude. 
Our virial mass estimates are based on single epoch line velocities and UV fluxes, and particularly for the most variable objects will carry a large uncertainty \citep[e.g.][]{Peterson,Vestergaard,Yong}.
In contrast to most other estimates reported here, the H$\beta$ broad line FWHM for XMMSL1\,J095336.4+161231 has been taken from a published fit using only a single Gaussian \citep{Alam} and at FWHM $= 3452$ km s$^{-1}$ is relatively narrow among our sources.
Black hole masses derived using the Virial method are significantly affected by host contamination for low luminosity (log\,$L_{5100\AA} \le 44.5$), nearby ($z \le 0.5$) AGN, leading to underestimation of the mass by of a few tenths of a dex according to \citet{Shen}. Our sources all have 5100\AA\ luminosities around that value, spanning $0.97 \le$ log\,$L_{5100\AA}$/44.5 $\le 1.02$, while only one of our source redshifts lies beyond $z=0.5$. From this we caution that contamination from the host will play some role but conclude that it should not dramatically alter the black hole mass results.
 
\begin{figure}
\begin{centering}
  \includegraphics[width=0.4\textwidth]{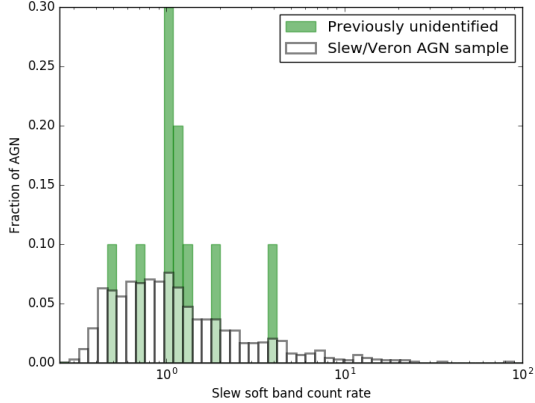}
  \caption{Comparison of Slew soft count rates for the secure Slew/Veron AGN and the AGN presented here.}
  \label{fig:CRhist}
\end{centering}
\end{figure}
\begin{figure}
  \begin{centering}
    \includegraphics[width=0.4\textwidth]{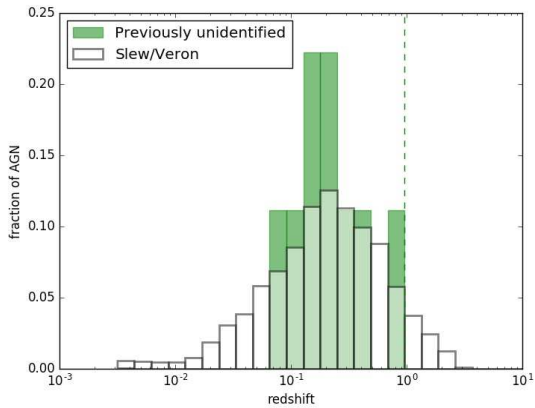}
\caption{Comparison of measured redshifts for the secure Slew/Veron AGN and the AGN presented here.}
\label{fig:zhist}
\end{centering}
\end{figure}

\subsection{Comparison with other XMM AGN samples}
\subsubsection{The full Slew Survey}
To examine our sample in the context of all Slew AGN, we compare with the secure AGN sample from the Slew Survey. The secure AGN sample, or Slew/Veron sample, is defined as all sources observed within XMMSL1-Delta-3 and {\it ROSAT} which
are contained in the \cite{Veron} catalogue of AGN \citep[for further details see][]{Saxton2011,Strotjohann}. The previously unidentified candidates now confirmed as AGN are those listed in Table \ref{tab:bhmass}. We compare their 0.2--2 keV soft band Slew count rates (see Table 1 of S11) with those of the Slew/Veron sample. We note that for one of our sources no soft band counts are recorded in the Slew survey, and for two sources there were two observations each, resulting in two values per source. Figure \ref{fig:CRhist} shows that the soft band count rates of the unidentified AGN cluster at around 1 ct s$^{-1}$, following the peak count rate of the Slew/Veron sample.

The redshifts are also consistent with the Slew/Veron redshift distribution, with a probability they are drawn from the same distribution of 0.62 using a K-S test. The two redshift distributions are compared in Figure \ref{fig:zhist}. 
Black hole mass estimates are not available for the full Slew/Veron sample, however we can compare our mass estimates with another XMM AGN sample which we describe in the following section.

\begin{figure}
  \begin{centering}
    \includegraphics[width=0.4\textwidth]{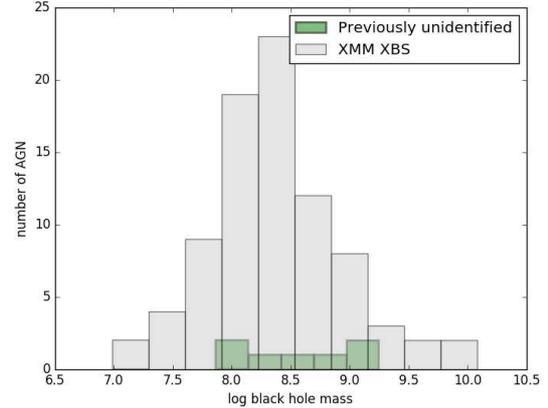}
\caption{Comparison of Virial black hole mass estimates for the XMM XBS AGN sample and the AGN presented here.}
\label{fig:bhmasshist}
\end{centering}
\end{figure}

\subsubsection{The XBS}
\cite{Caccianiga} analyse a complete, flux-limited ($S_{\rm X} > 7 \times 10^{-14}$ erg cm$^{-2}$ s$^{-1}$ at 0.5--4.5\,keV and 4.5--7.5\,keV) sample of Type I AGN from the {\it XMM-Newton} Bright Serendipitous Survey (XBS). They measure black hole masses from optical broad lines and UV continua for a subsample of 154 objects in a similar fashion to the method used here. Type I black hole masses in the XBS peak at 8$\times$10$^{8}$ M$_{\odot}$. These are consistent with the masses found here within uncertainties, while the mean is slightly larger than our Slew mean value of (5--6)$\times$10$^{8}$ M$_{\odot}$ (Figure \ref{fig:bhmasshist}). We note that the XBS extends to $z=2$ while the Slew Survey results presented here do not go above $z=1$ due to the accessibility of spectral lines in the chosen optical spectral range. The XBS values of Eddington ratio span 0.001--0.5, peaking at 0.1, where the Bolometric luminosity was computed from broadband SED fits. Our mean bolometric luminosity estimate lies within the range of the XBS Eddington ratios, and all but one of our estimates fall within the XBS range. 
Indications from these limited comparisons are that Slew Survey selected AGN are typically at lower redshift but show a similar range in black hole mass to AGN selected from the typically deeper fields in XMM pointed observations.

\subsection{Concluding remarks}
The use of {\it Swift} follow-up augmented by cross-correlation with optical and IR catalogues to distinguish likely AGN from likely Galactic stellar systems has been successful in producing a list of candidate AGN for optical spectroscopy.
We confirm the AGN nature and measure the redshifts of six XMM Slew Survey sources using optical spectroscopy, leading to 53 per cent of the candidate AGN from an originally unidentified Slew sample of S11 now confirmed. Among the proposed candidates, only 1--2 sources are not AGN. Most sources are of Type I, with one likely Type II system, in line with fractions found in large surveys. We identify one AGN which may be emitting close to the Eddington Limit. The measurable properties of these AGN fit well inside the envelopes of those of a constant sample of Veron AGN drawn from the XMM Slew Survey, and we find parallels with XBS AGN selected from pointed XMM observations.

Vast amounts of AGN candidates are expected from current and near-future X-ray surveys with e.g. ${\it Astrosat}$, ${\it eROSITA}$, ${\it SVOM}$, and from the stream of follow-ups of exotic events with large positional uncertainties. Through X-ray and then optical spectroscopic follow-up of shallow-survey X-ray sources that were initially difficult to identify, we are able to definitively classify about half the candidate AGN. Our data for $\sim$10--20 per cent of the candidate AGN suggest extraordinary properties worthy of further investigation, $\sim$5--10 per cent were shown to be Galactic in origin and the remainder were typical of known AGN populations. The effort required to obtain useful characterisation of initially unidentified sources is significant, yet it is key to population studies with wide-field X-ray and optical surveys.

\section{Acknowledgements}
We are grateful to the ING support astronomers O. Vaduvescu and L. Dominguez and WHT observer M. Cappellari for carrying out these service observations. The {\it Swift} BAT result was obtained with thanks to H. Krimm. Special thanks to Hannah Willett for contributions to the wider project during her Summer Undergraduate Research Experience with RLCS, funded by the University of Leicester. The authors wish to thank the anonymous referee for useful suggestions.
RLCS was supported by a Royal Society Dorothy Hodgkin Fellowship during the proposal and observation phases of this work; CW and KW acknowledge funding from STFC. SM acknowledges financial support by the Spanish Ministry of Economy and Competitiveness through grant AYA2016-76730-P, which is partly funded by the FEDER programme. BM acknowledges funding from the UK Space Agency.
The WHT and its service programme are operated on the island of La Palma by the Isaac Newton Group in the Spanish Observatorio del Roque de los Muchachos of the Instituto de Astrof\'isica de Canarias and we acknowledge data taken under service observing programme Sw2011b12. This work is based on observations collected at the European Organisation for Astronomical Research in the Southern hemisphere, Chile, programme ID 088.A-0628. The Pan-STARRS1 Surveys (PS1) and the PS1 public science archive have been made possible through contributions by the Institute for Astronomy, the University of Hawaii, the Pan-STARRS Project Office, the Max-Planck Society and its participating institutes, the Max Planck Institute for Astronomy, Heidelberg and the Max Planck Institute for Extraterrestrial Physics, Garching, The Johns Hopkins University, Durham University, the University of Edinburgh, the Queen's University Belfast, the Harvard-Smithsonian Center for Astrophysics, the Las Cumbres Observatory Global Telescope Network Incorporated, the National Central University of Taiwan, the Space Telescope Science Institute, the National Aeronautics and Space Administration under Grant No. NNX08AR22G issued through the Planetary Science Division of the NASA Science Mission Directorate, the National Science Foundation Grant No. AST-1238877, the University of Maryland, Eotvos Lorand University (ELTE), the Los Alamos National Laboratory, and the Gordon and Betty Moore Foundation. IRAF is distributed by the National Optical Astronomy Observatory, which is operated by the Association of Universities for Research in Astronomy (AURA) under a cooperative agreement with the National Science Foundation. This work made use of data supplied by the UK Swift Science Data Centre at the University of Leicester.

\bibliographystyle{mnras}
\bibliography{references}

\appendix
\section{Additional observational information} \label{sec:appendix}
\begin{figure*}
  \begin{centering}
    \includegraphics{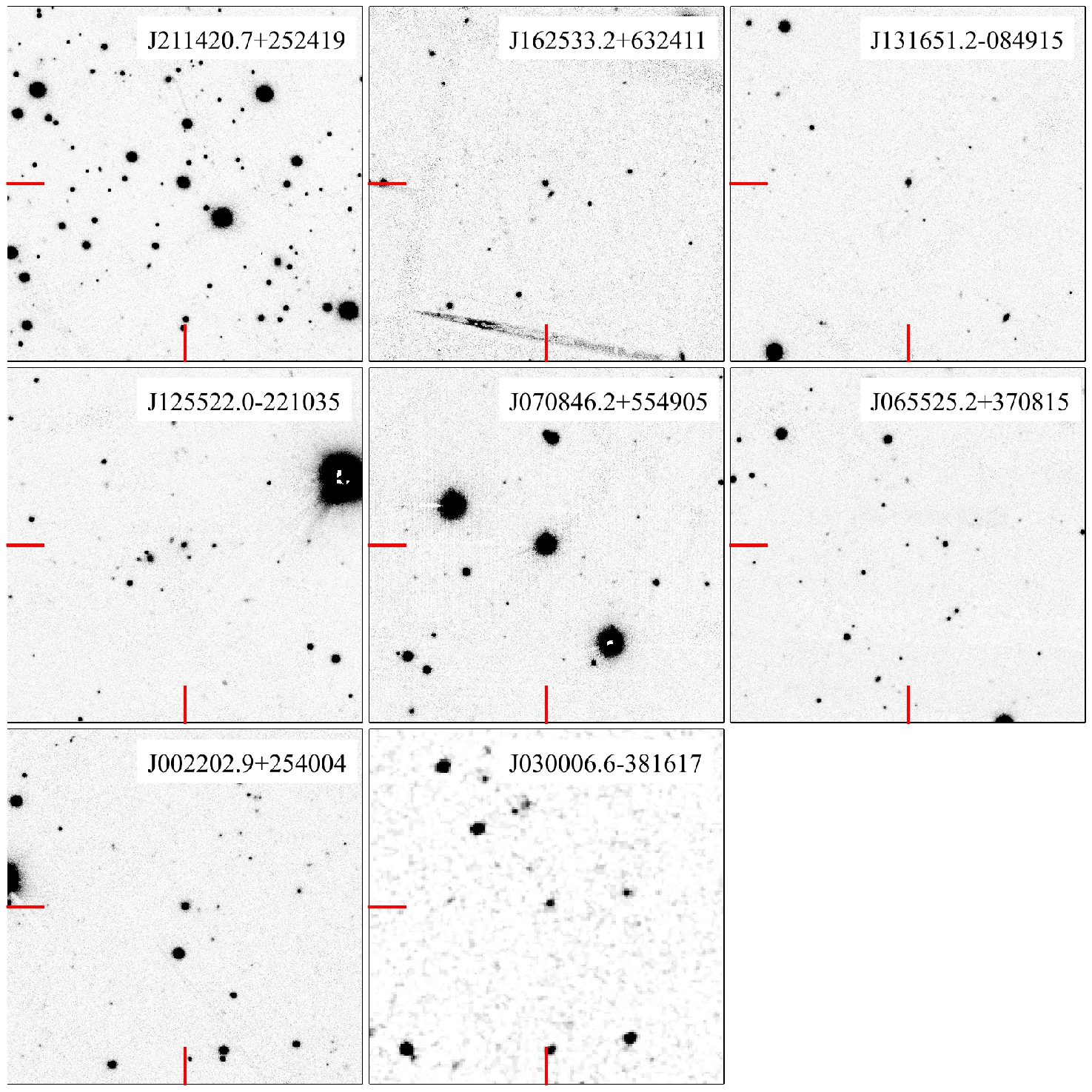}
\caption{Finding charts for each of the candidate AGN observed with WHT and VLT for this work. Each panel shows a $3\times2$ arc minute field of view, oriented north up and east left, centred on the source, where red tick marks point towards the source. Images are taken from from Pan-STARRS1 DR1 \citep{Chambers}, and are in $r$ band, except for XMMSL1\,J030006.6-381617, which is a DSS-2 Red image.}
\label{fig:fc}
\end{centering}
\end{figure*}

\bsp

\label{lastpage}

\end{document}